\documentclass[conference]{IEEEtran}
\IEEEoverridecommandlockouts

\usepackage{cite}
\usepackage{amssymb,amsfonts,amsthm,amsmath,mathrsfs}
\usepackage{algorithmic}
\usepackage{algorithm}
\usepackage{graphicx}
\usepackage{subcaption}
\usepackage{textcomp}
\usepackage{xcolor}
\def\BibTeX{{\rm B\kern-.05em{\sc i\kern-.025em b}\kern-.08em
    T\kern-.1667em\lower.7ex\hbox{E}\kern-.125emX}}

\usepackage{multirow, booktabs}
\usepackage{makecell}

\usepackage[colorlinks=true, linkcolor=blue, citecolor=blue]{hyperref}
\usepackage{url}
\usepackage{bm}
\usepackage{bbm}
\usepackage[super]{nth}

\newcommand{\var}{\mbox{var}}

\newcommand\dd{\mathrm{d}}

\DeclareMathOperator{\E}{\mathbb{E}}

\graphicspath{{./Figures/}}
  
\begin{document}

\title{Robust Analysis for Resilient AI System\\
\thanks{National Science Foundation DMS-2429324 and CMMI-2331985.}
}

\author{\IEEEauthorblockN{1\textsuperscript{st} Yu Wang}
\IEEEauthorblockA{\textit{Dept. of Mathematics and Statistics} \\
\textit{University of Massachusetts Amherst}\\
Amherst, U.S.A.\\
ywang6@umass.edu}
\and
\IEEEauthorblockN{2\textsuperscript{nd} Ran Jin}
\IEEEauthorblockA{\textit{Dept. of Industrial and Systems Engineering} \\
\textit{Virginia Tech}\\
Blacksburg, U.S.A. \\
jran5@vt.edu}
\and
\IEEEauthorblockN{3\textsuperscript{rd} Lulu Kang}
\IEEEauthorblockA{\textit{Dept. of Mathematics and Statistics} \\
\textit{University of Massachusetts Amherst}\\
Amherst, U.S.A.\\
lulukang@umass.edu}
}

\maketitle

\begin{abstract}
Operational hazards in Manufacturing Industrial Internet (MII) systems generate severe data outliers that cripple traditional statistical analysis. 
This paper proposes a novel robust regression method, DPD-Lasso, which integrates Density Power Divergence with Lasso regularization to analyze contaminated data from AI resilience experiments. 
We develop an efficient iterative algorithm to overcome previous computational bottlenecks. 
Applied to an MII testbed for Aerosol Jet Printing, DPD-Lasso provides reliable, stable performance on both clean and outlier-contaminated data, accurately quantifying hazard impacts. This work establishes robust regression as an essential tool for developing and validating resilient industrial AI systems.
\end{abstract}

\begin{IEEEkeywords}
AI system, density power divergence, resilience, robust regression, Lasso.
\end{IEEEkeywords}

\section{Introduction}

The integration of Artificial Intelligence (AI) into the Manufacturing Industrial Internet (MII) has become a paradigm for enhancing efficiency, enabling predictive maintenance, and optimizing complex processes. However, the deployment of AI systems in these critical, real-world environments exposes them to a multitude of operational hazards. 
As detailed in studies on AI resilience \cite{zeng2025fair}, these systems are vulnerable to disruptions across three interconnected layers: the data layer (e.g., sensor failures, distribution shifts), the AI pipeline layer (e.g., model singularity), and the cyber-physical layer (e.g., node failures, cyber-attacks). 
Such hazards frequently manifest as outliers and severely contaminate the data stream, leading to catastrophic performance degradation, erroneous predictions, and significant economic losses.

Consequently, investigating and ensuring the resilience of AI systems—defined as their ability to withstand, adapt to, and recover from disruptions—is paramount. 
This necessitates analytical frameworks capable of quantifying performance degradation and recovery under duress. 
As exemplified by the MII testbed for Aerosol Jet Printing (AJP) quality modeling \cite{zeng2025fair}, experiments designed to study AI resilience intentionally introduce a wide spectrum of these hazards (Table II in \cite{zeng2025fair} or Table \ref{tab:doe_factors1} in our paper), systematically generating data that is heavily contaminated with outliers. 
Traditional least squares regression, the most widely used method of statistical analysis, is notoriously sensitive to such anomalies. 
Its estimates can be disproportionately skewed by outliers, rendering it unsuitable for reliably analyzing resilience metrics or identifying the true impact of experimental factors in this context.

This vulnerability of classical methods underscores the critical need for robust regression techniques. 
Robust methods are specifically designed to limit the influence of outliers, providing stable and reliable parameter estimates even when the underlying data deviates from ideal assumptions. 
Therefore, to accurately analyze the experimental results from AI resilience studies and draw valid conclusions about system behavior under hazards, a robust analytical approach is not merely beneficial but essential.

The field of robust statistics has developed a wide array of estimators designed to be resistant to deviations from idealized model assumptions. 
Seminal approaches include M-estimators (e.g., Huber loss \cite{huber1964robust}), which bound the influence of outliers; R-estimators \cite{Heiler01011988}, based on ranks; and S-estimators \cite{10.1007/978-1-4615-7821-5_15}, which minimize a robust measure of scale of the residuals. 
These estimators have been extended to generalized linear models combined with variable selection or regularization methods, including \cite{Khan01122007}, \cite{ALFONS2016421}, \cite{chang2018robust}, \cite{loh2017statistical}, etc. 
While these methods offer significant improvements over OLS, they often come with their own limitations, such as low statistical efficiency under pure normality or a lack of conceptual unity in their design principles.

In recent decades, a powerful and elegant paradigm for deriving robust estimators has emerged from information theory: the use of divergences. 
Divergences, such as the Kullback-Leibler (KL) Divergence, measure the ``distance'' or discrepancy between two probability distributions. 
The fundamental connection to regression was established by recognizing that minimizing the OLS objective function is equivalent to minimizing the KL divergence between the assumed parametric model of the data and the true, underlying data-generating distribution. 
This insight opens the door to a generalized approach: replacing the KL divergence with a broader class of divergences that are inherently more robust to model misspecification.
Prominent examples of such robust divergences include:
\begin{itemize}
\item Density Power Divergence (DPD) by \cite{basu1998robust}, which introduces a tuning parameter that controls the trade-off between efficiency and robustness.
\item $\gamma$-Divergence by \cite{fujisawa2008robust}, which enjoys strong robustness properties, including immunity to outliers in the response variable.
\item Bregman Divergences \cite{BREGMAN1967200}, a broad class that generalizes the squared error loss.
\item Maximum Mean Discrepancy (MMD) \cite{sriperumbudur2011universality,sriperumbudur2010hilbert}, a distance on the space of probability measures. 
\end{itemize}
Using these divergences, many robust modeling approaches have been proposed. 
For example, \cite{ghosh2013robust} introduced a DPD minimization for linear regression;  \cite{chi2014robust} developed a robust penalized logistic regression algorithm based on a minimum distance criterion, which is a special case of DPD; \cite{alquier2023universal} built two estimators based on minimizing MMD that are both proven to be robust to Huber-type contamination. 
These divergence-based estimators are typically derived by minimizing the divergence between the empirical distribution of the data and the assumed model distribution. 
This process, often referred to as minimum divergence estimation, yields estimators that are highly robust while maintaining good efficiency. 
The tuning parameter inherent in most divergences provides a flexible mechanism for the practitioner to control the degree of robustness required for a specific application.

In this work, we employ a novel robust regression method based on the DPD \cite{basu1998robust} to analyze data from an AI resilience experiment. 
This method is particularly well-suited for this application as it is designed to maintain estimation efficiency under clean data conditions while offering strong resilience against outliers. 
Compared with \cite{ghosh2013robust}, we also add the lasso penalty for variable selection. 
\cite{alquier2023universal} criticized the DPD-based linear regression of \cite{ghosh2013robust} for the challenging computation in the optimization procedure. 
To overcome this limitation, we developed an iterative minimization procedure. 
In each iteration, the minimization becomes a weighted regression with $l_1$-regularization, which can be easily solved by the lasso method. 
The convergence of the algorithm is also guaranteed since it is essentially a block coordinate descent method for a convex optimization problem. 
For short, we name the proposed method DPD-Lasso. 

We demonstrate its application in evaluating the performance of an MII testbed in which the AI system supports the AJP quality modeling as its core computation task \cite{zeng2025fair}, showing how it provides more reliable insights into the system's response to controlled hazards compared to non-robust alternatives. 
Our analysis confirms that robust statistical approaches are a fundamental component in the toolkit for developing and validating resilient AI systems.

\section{Methodology}\label{sec:method}

In this section, we propose the loss function for estimating the unknown parameters of a linear regression model based on the density power divergence. 
To facilitate variable selection for applications with a large number of input variables, we also add $l_1$-norm regularization, i.e., the lasso penalty on the linear regression coefficients. 
To minimize the regularized loss function, we develop a minimization algorithm and a cross-validation procedure to select tuning parameters. 

\subsection{Regularized DPD Loss Function}\label{subsec:loss}

Consider a linear model
\begin{align}\label{eq:regression}
y = \bm x^\top \bm \beta + \epsilon,
\end{align}
where $\bm x  = (x_{1}, \ldots, x_{p})^\top$ is a vector of $p$ continuous predictor variables,
$y$ is the response value, $\bm \beta = (\beta_{1}, \ldots, \beta_{p})^{T}$ are regression parameters,
and the error term  $\epsilon$ is normally distributed with mean zero and variance $\sigma^{2}$ with the density function $\phi(\epsilon|0, \sigma^{2})$.
Suppose the data are centered so that the model in \eqref{eq:regression}
does not contain an intercept.
Assume this model is \textit{sparse} in that only $p_{0} < p$ predictor variables are \textit{active}.

The density power divergence \cite{basu1998robust}, denoted by $d_{\alpha}(f,g)$, measures the difference between two probability density functions $f$ and $g$. 
\begin{align}\nonumber
d_{\alpha}(g, f)&=\int f^{1+\alpha}(z)\dd z-\left(1+\frac{1}{\alpha}\right)\int g(z)f^{\alpha}(z)\dd z\\
&+\frac{1}{\alpha}\int g^{1+\alpha}(z)\dd z,  \label{eq:d_alpha}
\end{align}
where $\alpha>0$. When $\alpha=0$, the integral is not defined, and thus \cite{basu1998robust} defined $d_{0}(f,g)$ as the KL-divergence. 

Given the training data $\{\bm x_i, y_i\}_{i=1}^n$, we want to use $d_{\alpha}$ as the loss function to estimate the regression model \eqref{eq:regression}, instead of the square loss in OLS. 
We replace $f$ in $d_{\alpha}(f,g)$ by the density function of the random noise $\epsilon$, i.e., 
\[
f(\epsilon|0, \sigma^2)=\phi(\epsilon|0, \sigma^2)=\phi(y-\bm x^\top \bm \beta|0, \sigma^2).
\]
It can be directly derived that 
\begin{align*}
& \int f^{1+\alpha}(\epsilon|0, \sigma^2)\dd \epsilon=\int \left(\phi(\epsilon|0, \sigma^2)\right)^{1+\alpha}\dd \epsilon \\
& =\left(\frac{1}{\sqrt{2\pi}\sigma}\right)^{1+\alpha}\int \exp\left[-(1+\alpha)\frac{\epsilon^2}{2\sigma^2}\right]\dd \epsilon\\
&=\frac{1}{(\sqrt{2\pi}\sigma)^{\alpha}}\frac{1}{\sqrt{1+\alpha}}. 
\end{align*}
Regarding the choice of $g$, if we are going to use the parametric version, we can assume there are true parameter values $\bm \beta_0$ and $\sigma^2_0$. 
Then $g(\epsilon|0, \sigma_0^2)=\phi(\epsilon|0,\sigma_0^2)=\phi(y-\bm x^\top \bm \beta_0|0, \sigma_0^2)$. 
Also, $\int g^{1+\alpha}(z) \dd z$ does not depend on data, and thus can be omitted from the loss function. 
We only need derive $\int g(\epsilon|0,\sigma_0^2)f^{\alpha}(\epsilon|0, \sigma^2)\dd \epsilon$. 

\begin{align*}
& \int g(\epsilon|0,\sigma_0^2)f^{\alpha}(\epsilon|0, \sigma^2)\dd \epsilon = \int \phi(\epsilon|0,\sigma_0^2)\phi^{\alpha}(\epsilon|0, \sigma^2)\dd \epsilon\\
= &\mathbb{E}_{g}\left(f^{\alpha}(\epsilon | 0,\sigma^2)\right)\approx \frac{1}{n}\sum_{i=1}^n f^{\alpha}(\epsilon_i|0,\sigma^2) \\
= &\frac{1}{n}\left(\frac{1}{\sqrt{2\pi}\sigma}\right)^{\alpha}\sum_{i=1}^n\exp\left[-\frac{\alpha}{2\sigma^2} (y_i-\bm x_i^\top \bm \beta)^2\right]. 
\end{align*}
Since we do not know the true value $\bm \beta_0$ and $\sigma_0^2$, we use the sample mean of $f(\epsilon_i|0,\sigma^2)$ (with any given $\sigma^2$ and $\bm \beta$ value) where $\epsilon_i=y_i-\bm x_i^\top \bm \beta$ for $i=1,\ldots, n$ to approximate $\mathbb{E}_{g}\left(f^{\alpha}(\epsilon | 0,\sigma^2)\right)$. 
Using this approach, we can consider the following loss function $Q(\bm \beta, \sigma^2)$ to be minimized, 
\begin{align*}
&Q_{\alpha}(\bm \beta, \sigma^2)=\int f^{1+\alpha}(\epsilon|0,\sigma^2)\dd \epsilon\\
&-\left(1+\frac{1}{\alpha}\right)\int g(\epsilon|0,\sigma_0^2)f^{\alpha}(\epsilon|0,\sigma^2)\dd \epsilon\\
&=\frac{1}{(\sqrt{2\pi}\sigma)^{\alpha}}\frac{1}{\sqrt{1+\alpha}}\\
&-(1+\alpha^{-1})\frac{1}{n}\left(\frac{1}{\sqrt{2\pi}\sigma}\right)^{\alpha}\sum_{i=1}^n\exp\left[-\frac{\alpha}{2\sigma^2} (y_i-\bm x_i^\top \bm \beta)^2\right],
\end{align*}
which is further simplified by dropping the constant term
\begin{equation}\label{eq:Q}
Q_{\alpha}(\bm \beta, \sigma^2)=-\frac{1}{n}\sum_{i=1}^n \exp\left[-\frac{\alpha}{2\sigma^2} (y_i-\bm x_i^\top \bm \beta)^2\right].
\end{equation}
When $\alpha=1$, this becomes the $L_2E$ robust regression problem \cite{Scott01082001,https://doi.org/10.1002/wics.4} and the loss function is 
\[
\min_{\bm \beta, \sigma^2} -\frac{1}{n}\sum_{i=1}^n \exp\left[-\frac{1}{2\sigma^2} (y_i-\bm x_i^\top \bm \beta)^2\right].
\]
Minimizing $Q_{\alpha}(\bm \beta, \sigma^2)$ is equivalent to minimizing a log transformation of the sum of the exponential functions, i.e., the log-sum-exp trick
\begin{equation}\label{eq:minreg2}
\min_{\bm \beta, \sigma^2} -\log\left(\frac{1}{n}\sum_{i=1}^n \exp\left[-\frac{\alpha}{2\sigma^2} (y_i-\bm x_i^\top \bm \beta)^2\right]\right).
\end{equation}
The log-sum-exp trick is a common numerical technique to prevent numerical overflow or underflow. 
Therefore, we update the definition of $Q_{\alpha}$ to 
\begin{equation}\label{eq:logQ}
Q_{\alpha}(\bm \beta, \sigma^2)=-\log \left(\frac{1}{n}\sum_{i=1}^n \exp\left[-\frac{\alpha}{2\sigma^2}(y_i-\bm x_i^\top \bm \beta)^2\right]\right).
\end{equation}

For high-dimensional input problems, we can add the lasso penalty for variable selection. 
We are going to solve this minimization problem to estimate $\bm \beta$ and $\sigma^2$, 
\begin{equation}\label{eq:minreg}
\min_{\bm \beta, \sigma^2} -\log\left(\frac{1}{n}\sum_{i=1}^n \exp\left[-\frac{\alpha}{2\sigma^2} (y_i-\bm x_i^\top \bm \beta)^2\right]\right)+\lambda \|\bm \beta \|_1.
\end{equation}

\subsection{Minimization Algorithm}\label{subsec:alg}

For any positive $\alpha$, we derive the following minimization algorithm.
Denote $h_i(\bm \beta, \sigma^2)=-(y_i-\bm x_i^\top \bm \beta)^2/2\sigma^2$ which involves $(\bm x_i, y_i)$ and the parameters $(\bm \beta, \sigma^2)$. 
We can obtain the following approximation (omit $\sigma^2$ for simplicity) based on first-order Taylor expansion around $\bm \beta_0$,
\begin{align*}
& \log[n^{-1}\sum_{i=1}^n\exp(\alpha h_i(\bm \beta))]\approx C_0+ \\
& \sum_{i =1}^{n} \frac{\alpha\exp(\alpha h_{i}(\bm \beta_{0}))}{\sum_{j=1}^{n} \exp( \alpha h_{j}(\bm \beta_{0}) )} \nabla h_{i}(\bm \beta_{0})^\top (\bm \beta - \bm \beta_{0}).
\end{align*}
Also based on the first order Taylor approximation of $h_i(\bm \beta, \sigma^2)$ centered at $\beta^{(t)}$, which is the current $\bm \beta$ value at the $t$-th iteration of the minimization procedure,
\begin{align*}
& h_i(\bm \beta)\approx h_i(\bm \beta^{(t)})+\nabla h_{i}(\bm \beta^{(t)})(\bm \beta-\bm \beta^{(t)})\\
 \Rightarrow & \nabla h_{i}(\bm \beta^{(t)})(\bm \beta-\bm \beta^{(t)})\approx h_i(\bm \beta)- h_i(\bm \beta^{(t)}).
\end{align*}
Then the log of the sum of the exponential function is approximately, 
\begin{align*}
&\log\left[n^{-1}\sum_{i=1}^n\exp(\alpha h_i(\bm \beta))\right]\\
\approx & C_1+ \sum_{i =1}^{n} \frac{\alpha\exp(\alpha h_{i}(\bm \beta^{(t)}))}{\sum_{j=1}^{n} \exp( \alpha h_{j}(\bm \beta^{(t)}) )}h_i(\bm \beta).
\end{align*}
Here, both $C_0$ and $C_1$ are constants not involving the parameters. 
Therefore, the objective function can be rewritten into 
\begin{align}\label{eq:minregw}
\min_{\bm \beta,\sigma^2} & \quad \sum_{i=1}^n w_i^{(t)}(y_i-\bm x_i^\top \bm \beta)^2 +\lambda \|\bm \beta \|_1, \\\nonumber
& \text{where } w_i^{(t)}=\frac{\exp(\alpha h_i\left(\bm \beta^{(t)})\right)}{\sum_{j=1}^n \exp(\alpha h_j\left(\bm \beta^{(t)})\right)}. 
\end{align}
Note that this objective function is an approximation of the original objective \eqref{eq:minreg2}, omitting the irrelevant constant. 
We assume $\bm \beta^{(t)}$ should be close to $\bm \beta$ in the next step. 
Thus, this approximation is used in the iterative minimization of the original objective function \eqref{eq:minreg2}. 
We summarize the minimization procedure in Algorithm \ref{alg:reg}.

\begin{algorithm}
\caption{Algorithm for Robust Regression for High-Dimensional Input.}
\label{alg:reg}
\noindent
{\bf Step 0}: Choose the parameters $\alpha>0$, $\lambda>0$ and $\texttt{Tol}>0$. Here \texttt{Tol} is the threshold to measure convergence of the algorithm. 

{\bf Step 1}: Set $t=0$. Obtain an initial estimate of $\bm \beta^{(0)}$ using an existing regression approach such as least squares regression or least squares regression with shrinkage methods. Based on $\bm \beta^{(0)}$, the initial $\sigma_{0}^2$ can be the MLE of the regression model, i.e., $\sigma_0^2=\frac{1}{n}\sum_{i=1}^n (y_i-\bm x_i^\top \bm \beta^{(0)})^2$. 

{\bf Step 2}: Compute the weight $w_{i}^{(t)} = \frac{\exp(\alpha h_{i}(\bm \beta^{(t)}))}{\sum_{j=1}^{n} \exp(\alpha h_{j}(\bm \beta^{(t)}) )}$ for $i=1,\ldots, n$. 

{\bf Step 3}: Obtain $\bm \beta^{(t+1)}$ by minimizing \eqref{eq:minregw}, i.e.,
\begin{align*}
\bm \beta^{(t+1)} = \arg \min \sum_{i  =1}^{n} w_{i}^{(t)} ( y_{i} - \bm x_{i}^\top\bm \beta )^{2} + \lambda \|\bm \beta\|_1.
\end{align*}
Compute $\sigma^2_t$ by 
\[
\sigma^2_t = \frac{1}{n}\sum_{i=1}^n (y_i-\bm x_i^\top \bm \beta^{(t)})^2.
\]
{\bf Step 4}: Check convergence. If $\frac{\|\bm \beta^{(t)} - \bm \beta^{(t-1)}\|_2}{\|\bm \beta^{(t)}\|_2}\le \texttt{Tol}$ and $\|1-\frac{\sigma^2_t}{\sigma_{t-1}^2}|\leq \texttt{Tol}$, terminate the iteration and return $\bm \beta^{(t)}$ and $\sigma^2_t$. 
Otherwise, set $t \rightarrow t+1$ and return to  {\bf Step 2}.
\end{algorithm}

The weight $w_{i}$ is quite meaningful. 
When $\alpha = 0$, $w_{i}= 1, i  =1, \ldots, n$, the regression becomes Lasso regression. When $\alpha = 1$, it is $L_2E$ with lasso penalty. 
Therefore, $\alpha$ can control how much robustness (outlier resistance) is gained from the weight.
We can write 
\begin{align*}
\sum_{i = 1}^{n} w_{i}^{(t)} ( y_{i} - \bm x_{i}^\top\bm \beta )^{2} &= \sum_{i  =1}^{n}  ( \sqrt{w_{i}^{(t)}}y_{i} - \sqrt{w_{i}^{(t)}}\bm x_{i}^\top\bm \beta )^{2}  \\
&\equiv \sum_{i  =1}^{n}  ( y_{i}^{(t)} - (\bm x_{i}^{(t)})^\top\bm \beta )^{2},
\end{align*}
where $y_i^{(t)}=\sqrt{w_i^{(t)}}y_i$ and $\bm x_i^{(t)}=\sqrt{w_i^{(t)}}\bm x_i$.
So we can use the Lasso algorithm to solve the optimization in {\bf Step 3}. 
In our implementation, we use the \texttt{glmnet} function from the \texttt{R} package \texttt{glmnet} \cite{JSSv033i01} with the training data $\{y_i^{(t)}, \bm x_i^{(t)}\}_{i=1}^n$.
Other packages can be used in {\bf Step 3}, such as the Barzilai-Borwein method \cite{10.1093/imanum/8.1.141} implemented in the \texttt{R} package \texttt{BB} \cite{JSSv033i01}. 

Note that 
\[Q_{\alpha}(\bm \beta, \sigma^2)=-\log \left(\frac{1}{n}\sum_{i=1}^n \exp\left[-\frac{\alpha}{2\sigma^2}(y_i-\bm x_i^\top \bm \beta)^2\right]\right)\] 
is a convex function of $(\bm \beta, \sigma^2)$. 
Algorithm \ref{alg:reg} is essentially a version of the block coordinate descent method. 
Therefore, for the convex loss $Q(\bm \beta, \sigma^2)+\lambda \|\bm \beta\|_1$, Algorithm \ref{alg:reg} should converge based on the well-known theoretical properties of block coordinate descent \cite{beck2013convergence}. 

\section{Tuning Parameters}

We need to choose two tuning parameters for the DPD-based robust regression: parameter $\lambda$ for the weight of the $l_1$-norm penalty and parameter $\alpha$ in the DPD. 

The standard Lasso method for generalized linear models uses cross-validation to select the tuning parameter $\lambda$. 
Following the same idea, we also propose to use cross-validation to decide the $\lambda$ value. 
However, different from a conventional statistical modeling setting, we consider that there are outliers in the training data. 
Therefore, some more sophisticated sampling scheme other than completely random sampling should be used in order to make sure that each fold of the dataset is similarly distributed, especially in terms of proportions of outliers. 
Here, we propose to use stratified sampling based on the loss function values. 
Recall that the objective function takes the form of 
\[
Q_{\alpha}(\bm \beta, \sigma^2)+\lambda \|\bm \beta\|_1,
\]
where 
\begin{align*}
l_{i,\alpha}(\bm \beta, \sigma^2) &=\exp\left[-\frac{\alpha}{2\sigma^2}(y_i-\bm x_i^\top \bm \beta)^2\right], \\
\text{and } Q_{\alpha}(\bm \beta, \sigma^2)&=-\log \left(\frac{1}{n}\sum_{i=1}^nl_{i,\alpha}(\bm \beta, \sigma^2)\right).
\end{align*}
For now, we consider $\alpha$ is chosen and thus we drop $\alpha$ from $Q_{\alpha}(\bm \beta, \sigma^2)$ and $l_{i,\alpha}(\bm \beta, \sigma^2)$ to simplify notation. 
For any given $\lambda$ value, we can obtain the estimated parameter values $\hat{\bm \beta}$ using a robust regression approach, as well as the parameter $\sigma^2$. 
Then we can compute $l_i(\hat{\bm \beta}_{\lambda,}, \sigma^2_{\lambda})$ and sort them from smallest to largest, and denote the sorted $l_i$ scores by $l_{(i)}$. 
Suppose we are going to split the data into $K$ folds and $n$ is divisible by $K$. 
Algorithm \ref{alg:cv} is the $K$-fold cross-validation via stratified sampling based on $l$-score. 
\begin{algorithm}
\caption{Stratified Sampling Cross-Validation.}
\label{alg:cv}
\noindent
{\bf Step 0}: Specify the number of folds $K$. The tuning parameter $\alpha$ remains the same. For any given $\lambda$ value, obtain $(\hat{\bm \beta}_{\lambda}, \hat{\sigma}^2_{\lambda})$ from Algorithm \ref{alg:reg}. 

{\bf Step 1}: Compute the $l$-score, $l_i=\frac{(y_i-\bm x_i\hat{\bm \beta}_{\lambda})^2}{\hat{\sigma}_{\lambda}^2}$ and sort them from smallest to largest, $l_{(i)}(\hat{\bm \beta}_{\lambda}, \hat{\sigma}^2_{\lambda})$. 

{\bf Step 2}: Put the sorted $l_{(i)}$'s and their indices into $L=n/K$ sets $\mathcal{A}_{(l)}$. 

{\bf Step 3}: Stratified sampling to construct $K$ folds: from each of the $L$ sets, randomly choose one index $i$ to form a fold of $L$ indices. Remove the selected indices from all $\mathcal{A}_{(l)}$'s. Do this for $K$ times to construct $K$ folds of indices. 

{\bf Step 4}: Based on the $K$ folds of training data, conduct the cross-validation procedure. 
For linear regression, compute the cross-validation prediction error 
\[
CV= \frac{1}{n}\sum_{i=1}^n e_{i}^2, \text{ where } e_i=y_i-\bm x_i^\top \hat{\bm \beta}_{\lambda, -D(i)},
\]
where $\hat{\bm \beta}_{\lambda, -D(i)}$ is the $\bm \beta$ value estimated by all the $K-1$ folds of training data without the $D(i)$ fold that includes $(\bm x_i,y_i)$. 

{\bf Step 4:} Return the CV error. 
\end{algorithm}

The parameter $\alpha$ governs the trade-off between efficiency and robustness in the $d_{\alpha}$ divergence. 
\cite{basu1998robust} demonstrated that larger values of $\alpha$ diminish the influence of outliers on parameter estimates, though this conclusion was established within a single-parameter model. 
To investigate the effect of $\alpha$ on linear regression, we conducted a simulation study with the following setup.

Training data of size $n=100$ were generated with an input dimension of $p=2$. Each input vector $\bm x_i$ was drawn from a bivariate normal distribution with mean $\bm 0$ and a covariance matrix $\Sigma$ where $\Sigma_{i,j}=0.5^{|i-j|}$ and diagonal entries were 1. The mean of the output $y_i$ was set to $\mu_i=\E(y_i|\bm x_i)=\bm x_i^\top\bm \beta$ with linear coefficients $\bm \beta = (5, -5)$; no intercept was included. 
The observed response was $y_i=\mu_i+\epsilon_i$, where the error $\epsilon_i$ was independently sampled from $N(0, \sigma^2)$ and $\sigma^2$ was set to $0.05^2 \times s^2$, with $s^2$ being the sample variance of the $\mu_i$'s.

To study robustness, we contaminated $c\%$ of the training data, where the contamination rate $c\%$ was set to either $10\%$ or $20\%$. For these contaminated data entries, both the input and output were modified to be outliers.
The outlying inputs were independently sampled from $\text{Uniform}[-0.1,0.1]$ and $\text{Uniform}[1,2]$. 
The corresponding outlying outputs were generated as $y_i = x_{i,1}(-3) + x_{i,2}3$.

Contour plots of the least squares ($Q_{\alpha}$ with $\alpha=0$) loss function and the $Q_{\alpha}$ criterion (defined in \eqref{eq:Q}) are presented in Figures \ref{fig:contour} in the Appendix. 
These loss functions are expressed solely in terms of the parameter $\bm \beta$; note that since $Q_{\alpha}$ requires $\sigma^2$, we substituted the sample variance of the residuals. 
The first five sub-figures \ref{fig:c10_LS}--\ref{fig:c10_D2} correspond to contamination rates of $10\%$ and the rest five sub-figures \ref{fig:c20_LS}--\ref{fig:c20_D2} are for contamination rates of $20\%$. 
The red points indicate the true parameter values $\bm \beta=(5,-5)$, while the green points mark the minimizer of each loss function.

A comparison of the least squares loss with $Q_{\alpha}$ for $\alpha=0.25, 0.5, 1, 2$ reveals a clear pattern. 
The minimizer of the least squares loss deviates from the true parameter value, and this deviation worsens as the contamination rate increases from $10\%$ to $20\%$. 
In contrast, the minimizer of $Q_{\alpha}$ consistently recovers the true parameters across all values of $\alpha$ and contamination rates. 
This demonstrates the robustness of the $Q_{\alpha}$ loss function. 
Importantly, the value of $\alpha$ does not appear to significantly alter this robust behavior in this setting. 
While this is an empirical observation, we still recommend experimenting with different $\alpha$ values in practice.

\section{Numerical Experiments}\label{sec:num}

This section presents a comprehensive evaluation of the proposed robust regression estimator, DPD-Lasso. 
Through repeatable experiments, we assess the accuracy of the DPD-Lasso on three aspects: prediction, parameter estimation, and variable selection against established baseline methods in a high-dimensional setting with correlated predictors and adversarial contamination.
To mimic modern applications (e.g., genomics, econometrics), the simulations combine high dimensionality, multicollinearity, and carefully crafted outliers. 
For each contamination level, data generation, contamination, model fitting, and evaluation are repeated 100 times.

\subsection{Data Generation Process}

Data are generated from the sparse, high-dimensional linear model:
\begin{equation*}
y_i \;=\; \beta_0^{*} + \bm x_i^\top \bm \beta^{*} + \epsilon_i, \qquad i=1,\ldots,n,
\end{equation*}
where $y_i$ is the scalar response, $\bm x_i$ is a $p$-dimensional vector of predictors, $\beta_0^{*}$ is the true intercept, $\bm \beta^{*}$ is the true $p$-dimensional coefficient vector, and $\epsilon_i$ is the random error.

\paragraph{Model Parameters}
The training data consist of $n=1000$ observations with $p=50$ predictors. 
The true coefficient vector $\bm \beta^{*}$ is sparse, with $p_{\text{active}}=25$ nonzero elements, and the true intercept is $\beta_0^{*}=2.5$. 
The true coefficient vector is constructed to be sparse by randomly selecting the indices of the $p_{\text{active}}=25$ active predictors uniformly from $\{1,\ldots,p\}$. 
The nonzero coefficients are drawn from a uniform distribution $\text{Uniform}[0.5,1.5]$ to ensure a mixture of signal strengths. 
The predictor matrix follows a multivariate normal distribution with a first-order autoregressive (AR(1)) covariance structure:
\[
\mathbf{x}_i \sim \mathcal{N}(\mathbf{0},\,\boldsymbol{\Sigma}), \qquad
\Sigma_{jk}=\rho^{\,|j-k|}\ \text{ with }\ \rho=0.7.
\]
The error terms $\varepsilon_i$ are drawn independently from a normal distribution, $epsilon_i \sim \mathcal{N}(0,\sigma_{\epsilon}^{2})$.
The error variance $\sigma_{\epsilon}^{2}$ is carefully calibrated to achieve a target
signal-to-noise ratio (SNR) of $5$, where $\text{SNR} = \var(\bm x^\top \bm \beta^{*})/\var(\epsilon)$.

\subsection{Contamination Mechanism}
The contamination strategy is designed to be adversarial, creating outliers that have maximum impact. We study three levels of contamination, controlled by the parameter $c\% \in \{0\%,5\%,10\%\}$, which represents the fraction of contaminated observations. 
The process introduces both bad leverage points (outliers in the predictor space $\bm x$) and vertical outliers (outliers in the response $y$) to the same subset of observations, creating a structured and challenging form of data corruption.

For a fraction $c\%$ of the data, we first create bad leverage input points by perturbing a random subset of the predictor variables with large shifts, i.e., randomly adding or subtracting 20.
For the same data entries, we then introduce vertical outliers by shifting their response values by a large magnitude (adding or subtracting 20). 
The sign of the shift is opposite to the sign of their true residual. 

\subsection{Baseline Methods}
We compare the DPD-Lasso with the following baseline methods. 
\begin{itemize}
\item \textbf{Lasso}: 
The $\ell_1$-penalized least-squares estimator, implemented using the \texttt{glmnet} \cite{R-glmnet} package in \texttt{R}. 
The regularization parameter $\lambda$ is chosen via 20-fold cross-validation to minimize prediction error.

\item \textbf{LAD-Lasso}
The Least Absolute Deviations (LAD) Lasso combines an $\ell_1$ penalty for sparsity with an $\ell_1$ loss for regression, providing robustness to vertical outliers (heavy-tailed errors or large shifts in $y$).
We use the implementation from the \texttt{R} package \texttt{MTE} \cite{LiQin2025MTE}, particularly the \texttt{LADlasso} function. 
\item \textbf{SparseLTS} The Sparse Least Trimmed Squares estimator, implemented via the \texttt{sparseLTS} function from the \texttt{robustHD} package \cite{Alfons2021robustHD} in \texttt{R}.
This method combines an $\ell_1$ penalty with the LTS objective. 
\end{itemize}

\subsection{Performance Metrics}
We use three complementary metrics to assess predictive power, estimation accuracy, and variable-selection performance.
\begin{enumerate}
\item \textbf{Root Mean Squared Prediction Error (RMSPE)} measures the model's predictive accuracy on the test set. It is defined by
\[
\mathrm{RMSPE}= \sqrt{\frac{1}{n_{\text{test}}}\sum_{i=1}^{n_{\text{test}}}
\bigl(y_{i,\text{test}} - \bm x_{i,\text{test}}^\top\hat{\bm \beta}\bigr)^2}.
\]
\item \textbf{$L_2$ Estimation Error} measures the fidelity of the estimated coefficient vector to the true vector $\boldsymbol{\beta}^{*}$:
\[L_2\ \mathrm{Error}= \|\hat{\bm \beta}-\bm \beta^{*}\|_{2}.\]
\item \textbf{Variable Selection Error ($\gamma$)} quantifies the errors in identifying the correct set of active predictors.
It is defined as the total number of false positives FP (inactive variables incorrectly selected) and false negatives FN (active variables incorrectly omitted): 
\[\gamma= \text{FP}+\text{FN}.\]
\end{enumerate}

\subsection{Results and Discussion}

\begin{figure}
    \centering
    \includegraphics[width=1\linewidth]{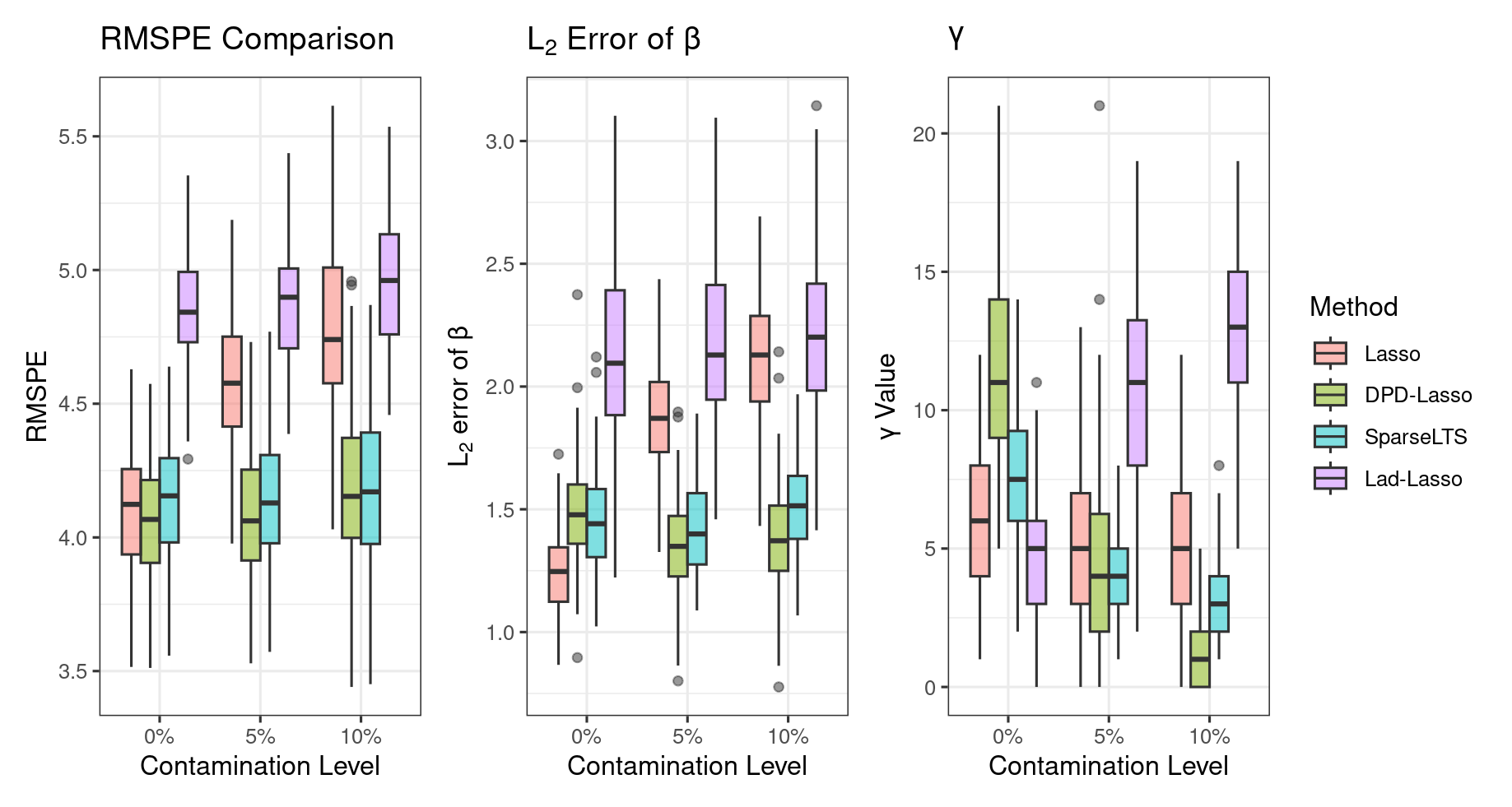}
    \caption{Box plots of RMSPE, $L_2$ Error of $\beta$, and variable selection error $\gamma$ of 100 simulations of four different methods.} \label{fig:simulation}
    \label{fig:placeholder}
\end{figure}

The boxplots in Figure \ref{fig:simulation} provide a comparative analysis of the methods' performance. 
In the zero-outlier scenario ($c\%=0\%$), which serves as a baseline for assessing statistical efficiency under ideal conditions, we observe that Lasso, DPD-Lasso, and SparseLTS achieve comparable performance in terms of RMSPE, while Lad-Lasso performs significantly worse. 
As expected, the non-robust Lasso demonstrates high efficiency across all three metrics—RMSPE, $L_2$ error, and variable selection accuracy ($\gamma$). 
The proposed DPD-Lasso method achieves the smallest RMSPE and a slightly higher $L_2$ error than both Lasso and SparseLTS. 
However, DPD-Lasso exhibits a substantially larger $\gamma$ value, indicating that the cross-validation procedure for selecting the tuning parameter $\lambda$ prioritized prediction accuracy at the expense of variable selection. 
This suggests a need for an improved $\lambda$ selection method to better balance these objectives. 
SparseLTS shows slightly worse RMSPE than DPD-Lasso, similar $L_2$ error, but superior performance in $\gamma$. 
Notably, both robust methods (DPD-Lasso and SparseLTS) are outperformed by Lasso on the $\gamma$ metric in this uncontaminated setting.

The introduction of adversarial contamination reveals a stark contrast in robustness. 
The performance of the non-robust Lasso deteriorates dramatically across all metrics. 
Our contamination mechanism, which injects both bad leverage points and large vertical outliers, creates spurious correlations that severely bias the estimated regression hyperplane away from the true model. 
This results in a sharp increase in prediction error (RMSPE), parameter estimation error ($L_2$), and variable selection error ($\gamma$). 
Interestingly, Lad-Lasso, whose $L_1$-loss offers some inherent robustness to vertical outliers, performs even worse than Lasso in this setting. 
This is because Lad-Lasso remains highly susceptible to bad leverage points in the design matrix, which corrupt the model fit and degrade its variable selection capabilities.

In contrast, both DPD-Lasso and SparseLTS demonstrate remarkable stability under contamination levels up to 10\%, maintaining strong performance across prediction, estimation, and variable selection. 
Compared to SparseLTS, DPD-Lasso exhibits a small but consistent improvement in prediction and estimation accuracy, as evidenced by lower medians and tighter interquartile ranges in the corresponding boxplots. 
Both methods excel at variable selection when data are contaminated, with the marginal advantage of DPD-Lasso likely attributable to its more robust tuning procedure. 
In summary, the proposed DPD-Lasso method achieves a highly desirable combination of properties: it matches the efficiency of non-robust methods on clean data while providing strong, reliable performance in the presence of adversarial outliers.

\section{Case Study}\label{sec:real}

The experimental data for this case study were generated from an MII testbed in which the AI system supports the AJP quality modeling as its core computation task \cite{zeng2025fair}. 
The dataset for the AI system is a real-world Aerosol Jet Printing (AJP) dataset with a multivariate time series (MTS) classification DNN pipeline. 
More details about the AJP dataset can be found in \cite{zeng2025fair}.
The AJP dataset consisted of 95 samples, each containing time series data from six in situ process variables—atomizer gas flow, sheath gas flow, current, nozzle X-coordinate, nozzle Y-coordinate, and nozzle vibration—alongside a binary quality label (conforming/nonconforming) based on measured circuit resistance. 
To increase complexity and better simulate a realistic industrial setting, four synthetic time series variables with negligible predictive power were added, resulting in a final set of ten input features. 
This data was then algorithmically augmented to emulate the operational data from five distinct virtual AJP machines, creating a robust testbed for evaluating AI resilience.

The primary goal of experimenting with the MII testbed was to validate the proposed resilience framework by subjecting the AI system to a comprehensive suite of controlled hazards across its data, AI pipeline, and cyber-physical layers. 
A full factorial experimental design was employed, systematically varying five root cause factors—such as the percentage of failed sensors, signal-to-noise ratio, input distribution shift, class imbalance, pipeline singularity, and node and communication failures—to create 1,458 distinct hazard scenarios. 
Due to the fact that setting the node and communication failures to any non-zero percentage would lead to complete failure of the AI system, we removed these two settings and the corresponding data entries from this study. 
Table \ref{tab:doe_factors1} shows the five root causes for hazards. 
Besides the root cause factors, the AI system also contains five fog machines (denoted by F).
Depending on which fog machine is used, three different DNN models are available for the classification task. 
Table \ref{tab:doe_factors2} shows the fog machine factor and the choice of DNN models factor. 

The objective of the analysis was to quantitatively assess the system's performance degradation under these adversarial conditions and, subsequently, to evaluate the efficacy of the proposed Multimodal Multi-head Self Latent Attention (MMSLA) model in diagnosing the root causes and enabling effective mitigation strategies to recover AI performance.
Therefore, we focus on two outcome variables: Inference Time and F1 score. 
Inference Time is the time spent on making the classification task on the AJP test dataset, and the F1 score measures the corresponding classification accuracy.

\begin{figure} 
\centering %
\begin{subfigure}[b]{0.45\textwidth}
\centering
\includegraphics[width=\textwidth]{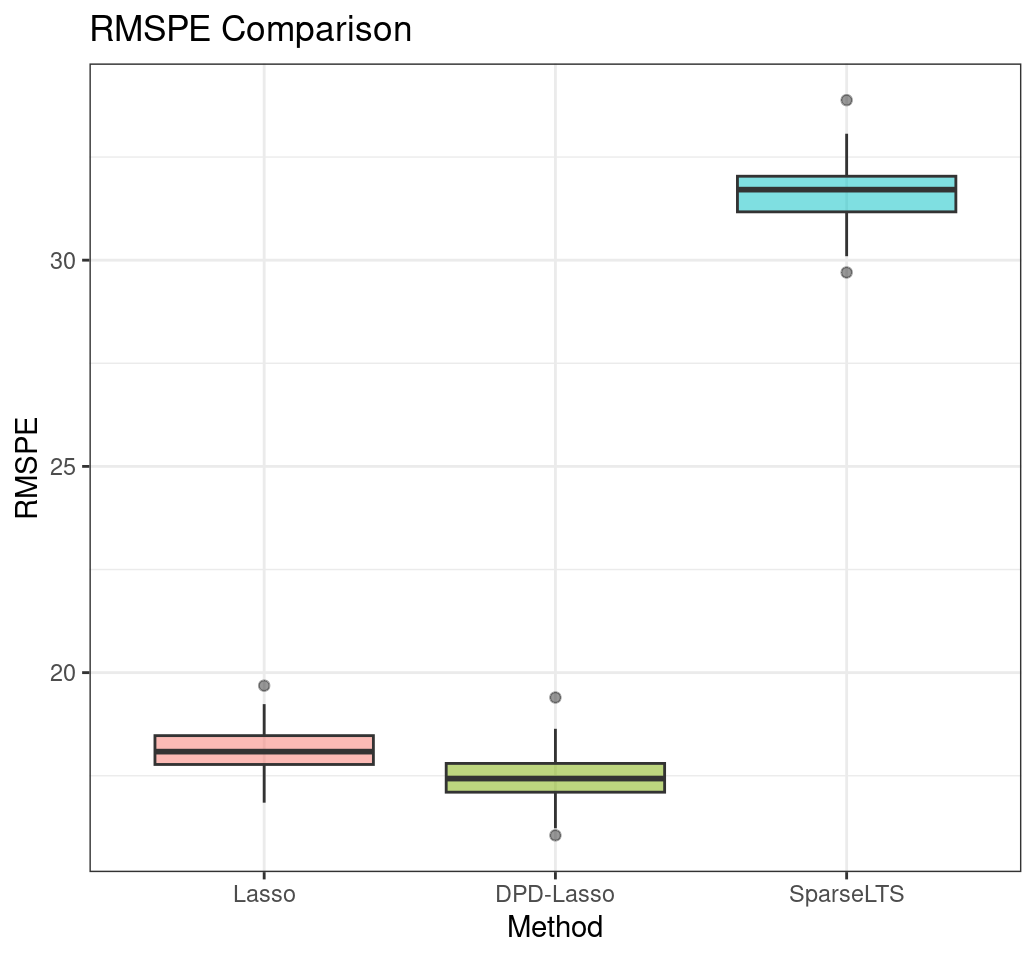} 
\caption{Comparison of RMSPE for Inference Time Prediction.}
\label{fig:ls_surface}
\end{subfigure}
\begin{subfigure}[b]{0.45\textwidth}
\centering
\includegraphics[width=\textwidth]{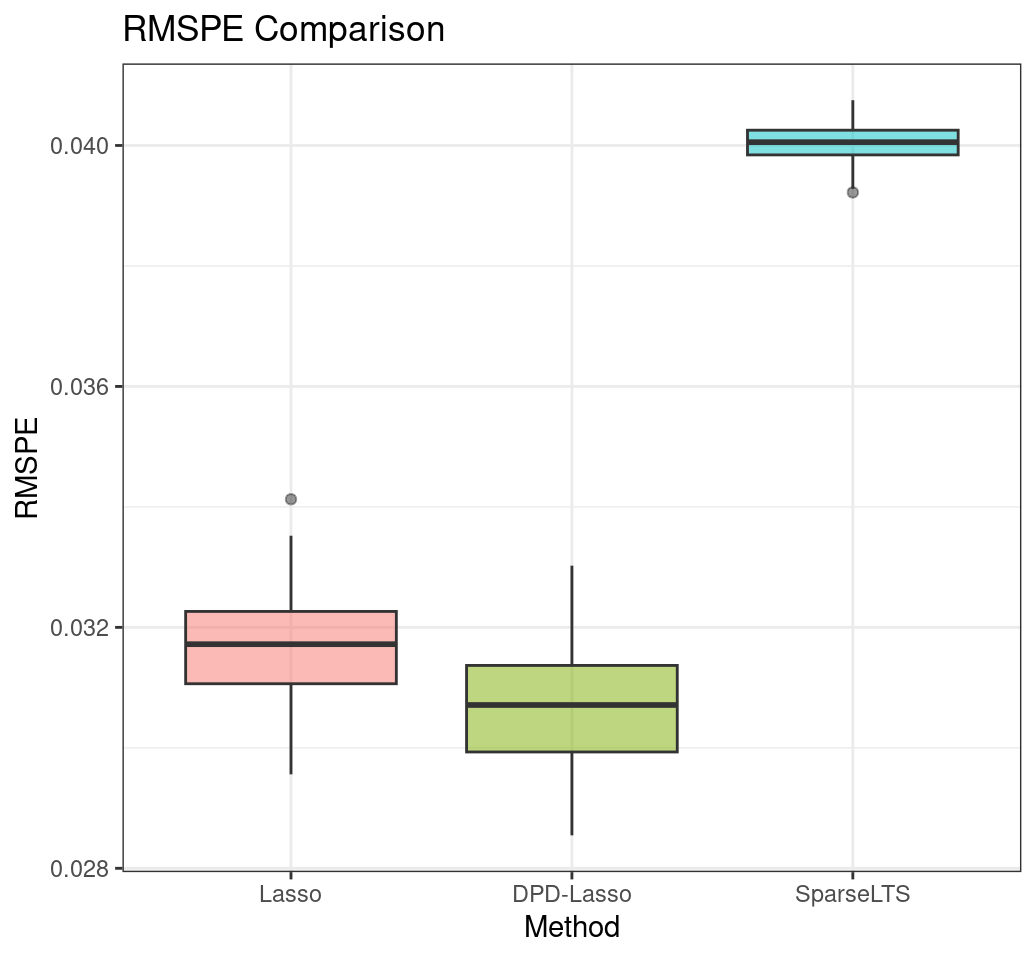}
\caption{Comparison of RMSPE for $\text{F}_1$ Score Prediction.}
\label{fig:d_alpha_surface}
\end{subfigure}
\caption{Comparison of RMSPE for Different Prediction Objects.}
\label{fig:surfaces_comparison}
\end{figure}

The dataset from the experiment on the MII testbed contains $n=14901$ observations and $p=20$ predictors, including all dummy variables representing the factors in Table \ref{tab:doe_factors1} and \ref{tab:doe_factors2}. 
We employed a repeated random sub-sampling methodology to ensure a robust evaluation. 
In this framework, the dataset was repeatedly partitioned by randomly selecting 70\% of observations, which amounts to $\lfloor 4901 \times 0.70 \rfloor = 10430$ samples, to be the training set in each run.  
The remaining 30\% of the data served as the test set. 
This entire process of random splitting, model training, and evaluation was replicated 100 times. 
Since this is a real dataset, we could not compare $L_2$ Error and variable selection. 
Figure \ref{fig:surfaces_comparison} shows the comparison of RMSPE for both Inference Time and F1 score. 
Due to the poor performance of Lad-Lasso, we did not include this method here. 
The proposed DPD-Lasso significantly outperforms Lasso and SparseLTS.
Therefore, DPD-Lasso is a suitable method for analyzing the experiments from the MII testbed, which can lead to better diagnostics, monitoring, and optimization of the resilient AI systems.

\begin{table}[htb] 
\centering
\caption{Experimental Design - Factors for Root Cause}
\label{tab:doe_factors1}
\begin{tabular}{@{}l l ccc@{}}
\toprule
\textbf{Factors} & \textbf{Definitions} & \textbf{0} & \textbf{1} & \textbf{2} \\
\midrule
A & \makecell{\% of Sensors Contaminated \\or Failed} & 0\% & 10\% & 20\% \\\hline
B & S/N ratio & High & -- & Low \\\hline
C & \makecell{Distribution Changes of\\ Input Variables (KL Divergence)} & 0 & Medium & High \\\hline
D & \makecell{Balanceness\\ (Binary Classification)} & 40-60 & 25-75 & 10-90 \\\hline
E & \% of Singular Pipelines (E)* & 0 & 1 & 2 \\
\bottomrule
\end{tabular}
\end{table}

\begin{table}[htb]
\centering
\caption{Experimental Design - Other Factors}
\label{tab:doe_factors2}
\begin{tabular}{l|l|l}\toprule
\textbf{Factors} & \textbf{Definitions} & Levels \\
\midrule
F & Machines  &  5-level, $M_1\sim M_5$ \\
$Z_1$ & Models used by $M_1$ & 3-level \\
$Z_2$ & Models used by $M_2$ & 3-level \\
$Z_3$ & Models used by $M_3$ & 3-level \\
$Z_4$ & Models used by $M_4$ & 3-level \\
$Z_5$ & Models used by $M_5$ & 3-level \\
\bottomrule
\end{tabular}
\end{table}

\section{Conclusion}

In this work, we addressed the critical challenge of analyzing contaminated data from AI resilience experiments in the Manufacturing Industrial Internet (MII). The proposed DPD-Lasso method effectively combines the robustness of Density Power Divergence with the variable selection capabilities of Lasso regularization, providing a powerful tool for reliable statistical inference in the presence of outliers. The developed iterative algorithm ensures computationally feasible and convergent optimization, overcoming a key limitation of previous DPD-based approaches.

Applied to a real-world MII testbed for Aerosol Jet Printing, DPD-Lasso successfully quantified the impact of diverse hazards across data, AI pipeline, and cyber-physical layers. Our analysis confirmed its dual strengths: it achieved efficiency comparable to non-robust methods like ordinary least squares under ideal conditions and demonstrated strong resilience against adversarial outliers, outperforming other methods. This reliability is paramount for drawing valid conclusions about system behavior under duress and for informing effective mitigation strategies.

This research underscores that robust statistical frameworks are not merely supplementary but fundamental to the entire lifecycle of resilient AI systems, from diagnosis and validation to recovery. Future work will focus on extending the DPD-Lasso framework to handle other types of generalized linear models and exploring online mitigation strategies that leverage real-time robust analytics for autonomous system recovery.

\section*{Acknowledgment}

The work of Yu Wang and Dr. Lulu Kang was supported by NSF DMS-2429324. The work of Dr. Ran Jin was supported by CMMI-2331985. 

\section*{Appendix}

This Appendix contains the contour plots of different loss functions for $Q_{\alpha}$ where $\alpha=0, 0.25, 0.5, 1, 2$ with $c\%=10\%$ or $20\%$ contaminated data by outliers. 

\onecolumn

\twocolumn

\onecolumn
\begin{figure}
\centering
\begin{subfigure}[b]{0.3\textwidth}
\centering
\includegraphics[width=\textwidth]{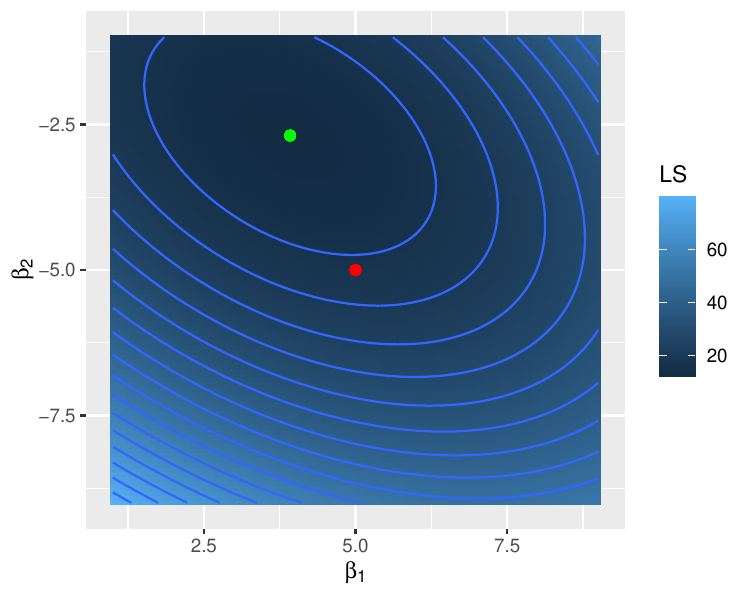}
\caption{Least square loss with 10\% outliers.}\label{fig:c10_LS}
\end{subfigure}
\begin{subfigure}[b]{0.3\textwidth}
\centering
\includegraphics[width=\textwidth]{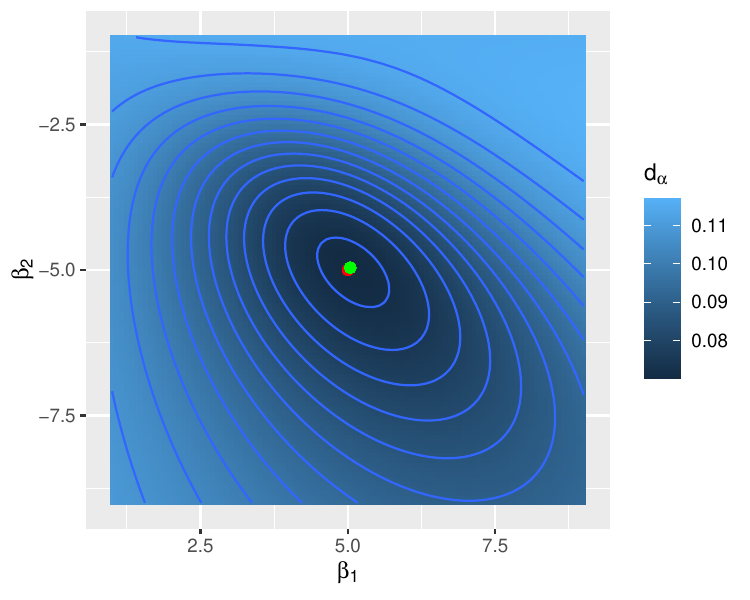}
\caption{$Q_{\alpha=0.25}$ with 10\% outliers.}\label{fig:c10_D025}
\end{subfigure}
\begin{subfigure}[b]{0.3\textwidth}
\centering
\includegraphics[width=\textwidth]{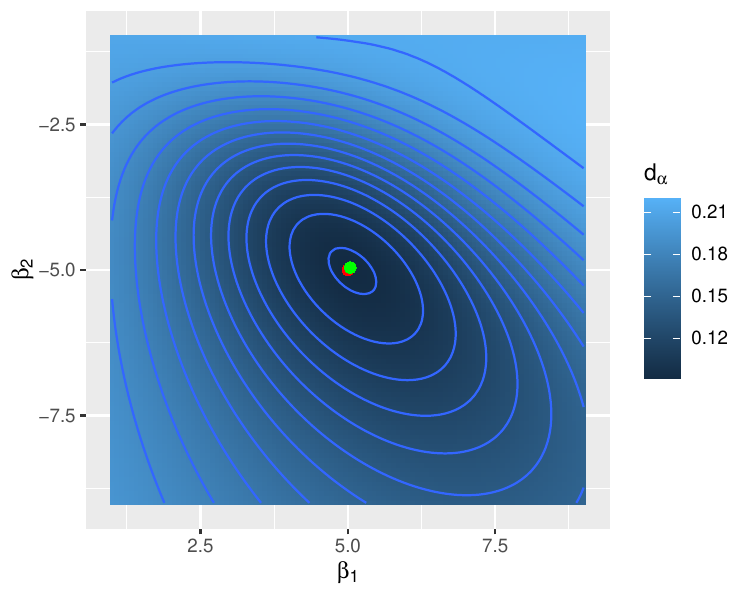}
\caption{$Q_{\alpha=0.5}$ with 10\% outliers.}\label{fig:c10_D05}
\end{subfigure}
\begin{subfigure}[b]{0.3\textwidth}
\centering
\includegraphics[width=\textwidth]{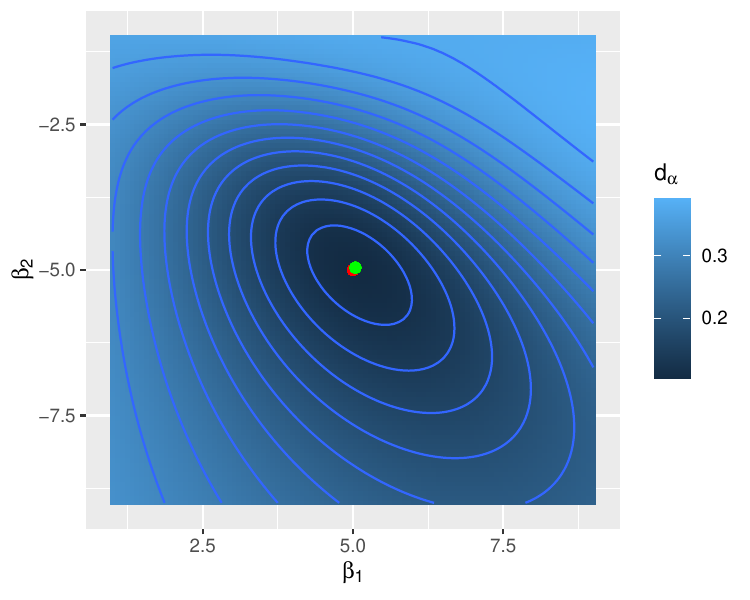}
\caption{$Q_{\alpha=1}$ with 10\% outliers.}\label{fig:c10_D1}
\end{subfigure}
\begin{subfigure}[b]{0.3\textwidth}
\centering
\includegraphics[width=\textwidth]{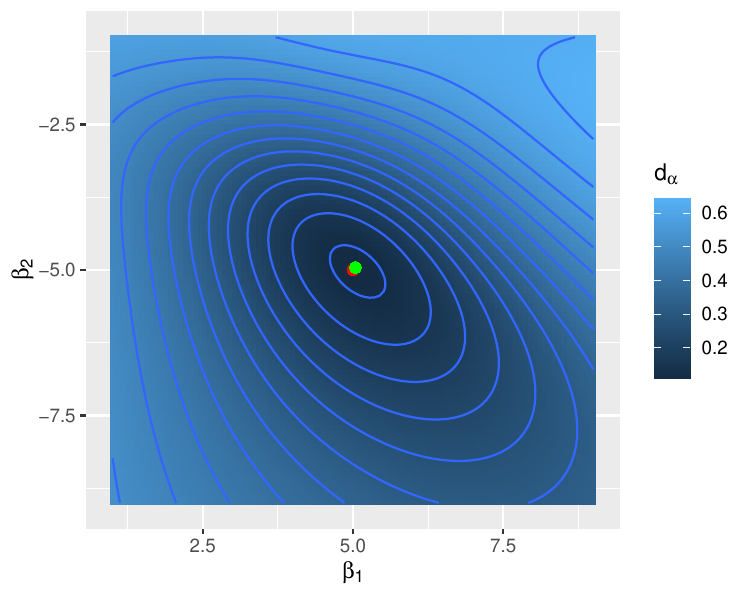}
\caption{$Q_{\alpha=2}$ with 10\% outliers.}\label{fig:c10_D2}
\end{subfigure}
\begin{subfigure}[b]{0.3\textwidth}
\centering
\includegraphics[width=\textwidth]{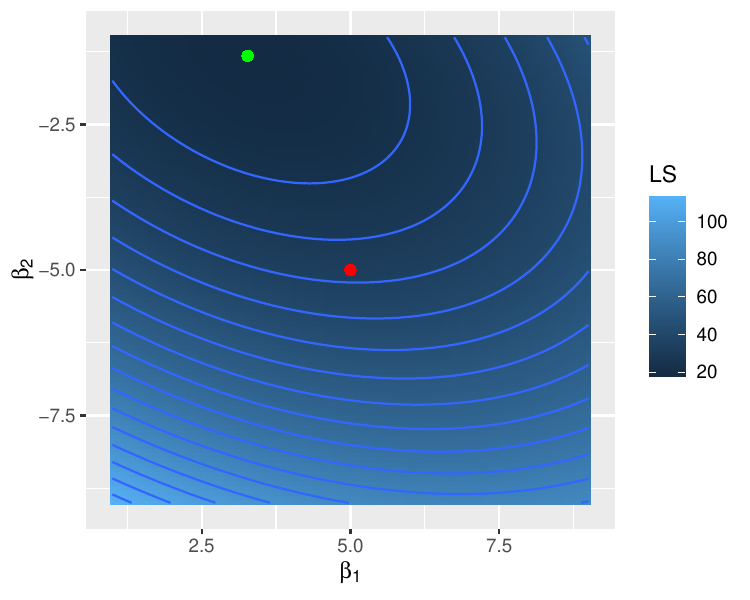}
\caption{Least square loss with 20\% outliers.}\label{fig:c20_LS}
\end{subfigure}
\begin{subfigure}[b]{0.3\textwidth}
\centering
\includegraphics[width=\textwidth]{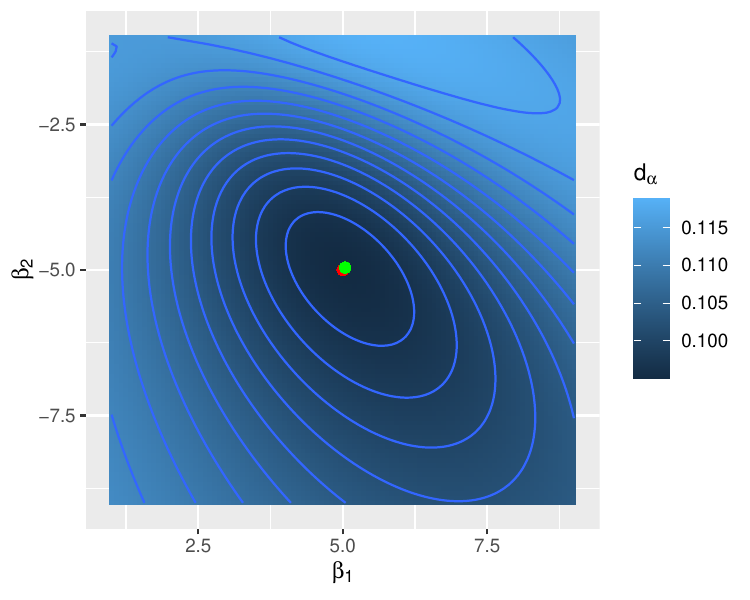}
\caption{$Q_{\alpha=0.25}$ with 20\% outliers.}\label{fig:c20_D025}
\end{subfigure}
\begin{subfigure}[b]{0.3\textwidth}
\centering
\includegraphics[width=\textwidth]{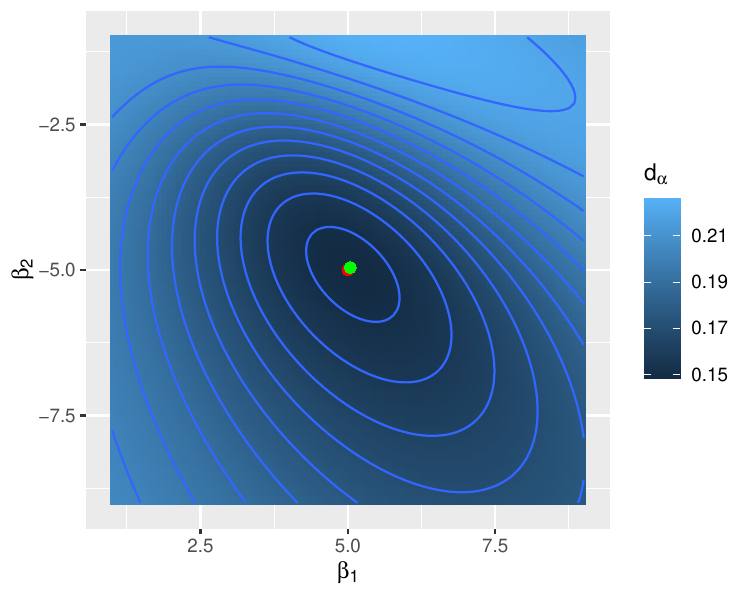}
\caption{$Q_{\alpha=0.5}$ with 20\% outliers.}\label{fig:c20_D05}
\end{subfigure}
\begin{subfigure}[b]{0.3\textwidth}
\centering
\includegraphics[width=\textwidth]{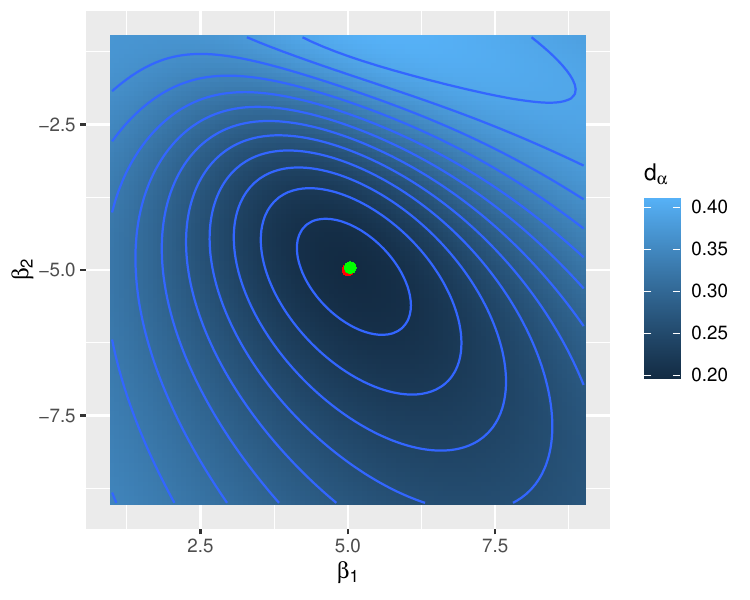}
\caption{$Q_{\alpha=1}$ with 20\% outliers.}\label{fig:c20_D1}
\end{subfigure}
\begin{subfigure}[b]{0.3\textwidth}
\centering
\includegraphics[width=\textwidth]{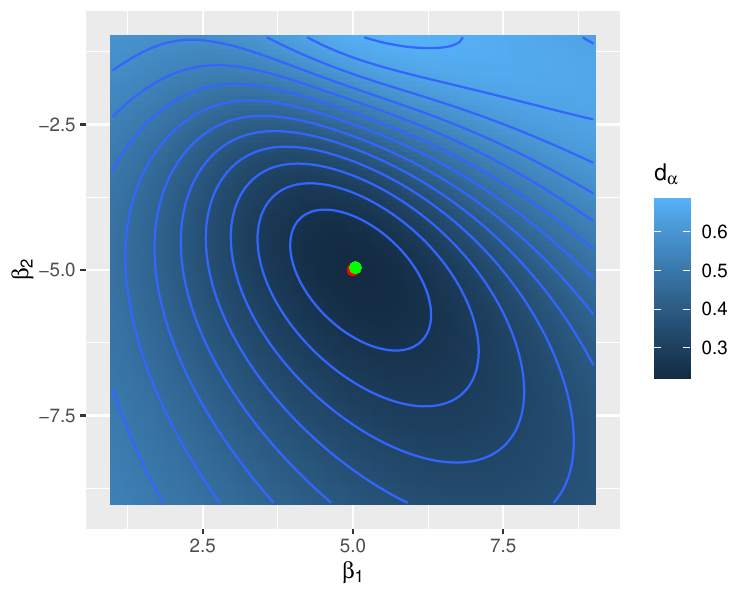}
\caption{$Q_{\alpha=2}$ with 20\% outliers.}\label{fig:c20_D2}
\end{subfigure}
\caption{Panel of 10 contour plots of least square loss, $Q_{\alpha=0.25}$, $Q_{\alpha=0.5}$, $Q_{\alpha=1}$, and $Q_{\alpha=2}$ with $10\%$ and $20\%$ outliers.}
\label{fig:contour}
\end{figure}
\twocolumn

\bibliographystyle{IEEEtran}
\bibliography{IEEEabrv,ref}

\end{document}